\documentclass[12pt,thmsa]{article}
\usepackage{sw20lart}

\input{tcilatex}
\begin{document}

\title{Tribimaximal mixing and leptogenesis in a seesaw model}
\author{Riazuddin \\
Centre for Advanced Mathematics and Physics,\\
National University of Science and Technology, Rawalpindi, Pakistan \\
and \\
National Centre for Physics, Quaid-i-Azam University,\\
Islamabad, Pakistan.}
\maketitle

\begin{abstract}
It is pointed out that if a neutrino mass matrix for right-handed neutrinos
in seesaw mechanism has $\mu -\tau $ symmetry and total lepton number $L_{e}$%
+$L_{\mu }$+$L_{\tau }$ remains constant (not zero), exact tribimaximal
neutrino mixing in this sector is produced. The same follows for the
effective Majorana light neutrino mass matrix provided that Yukawa couplings
(multiplied by the corresponding Higgs vacuum expectation values) in Dirac
mass matrix satisfy some constraints which in general implies zero
leptongenesis asymmetry. However this can be avoided when two of the heavy
right-hand neutrinos [the third one is irrelevant when $\mu -\tau $ symmetry
is assumed] are (nearly) degenerate.
\end{abstract}

In a previous paper [1] leptogenesis was studied in a seesaw model with $\mu
-\tau $ symmetry for $SU_{L}(2)$-singlet right handed neutrinos in the gauge
group $SU_{L}(2)\times U_{e}(1)\times U_{\mu -\tau }(1)$.

In addition to the usual fermions and Higgs, there are $SU_{L}(2)$- singlet
right handed neutrinos $N_{R}^{i}(i=e,\mu ,\tau )$ and the Higgs with
quantum numbers given below 
\begin{eqnarray}
L_{e} &:&(2,-1,0),\text{ }\phi ^{(1)}:(2,-1,0),\text{ }N_{R}^{e}\text{ :}%
(1,-1,1)  \nonumber \\
e_{R} &:&(1,-2,0)  \nonumber \\
L_{\mu -\tau } &:&(2,0,-1),\text{ }\phi ^{(2)}:(2,0,-1),\text{ }N_{R}^{\mu
,\tau }\text{ :}(1,1,-1)  \nonumber \\
\mu _{R},\tau _{R} &:&(1,0,-2)  \nonumber \\
\Sigma &:&(1,0,0)  \nonumber \\
\Sigma ^{\prime } &:&(1,2,-2)  \label{pp2}
\end{eqnarray}

The Yukawa couplings of neutrinos with Higgs, using $\mu -\tau $ symmetry
for right-handed neutrinos only, is given by [suppressing subscripts $L$ and 
$R$] 
\begin{eqnarray}
\mathcal{L}_{Y} &=&\left. 
\begin{array}{c}
h_{11}\overline{L}_{e}N_{e}\phi ^{(2)}+[h_{22}\overline{L}_{\mu }(N_{\mu
}+N_{\tau })+h_{32}\overline{L}_{\tau }(N_{\mu }+N_{\tau })]\phi ^{(1)} \\ 
+hc+f_{11}N_{e}^{T}CN_{e}\Sigma ^{\prime }+f_{12}N_{e}^{T}C(N_{\mu }+N_{\tau
})\Sigma +hc \\ 
+f_{22}[(N_{\mu }^{T}CN_{\mu }+N_{\tau }^{T}CN_{\tau })+f_{23}(N_{\mu
}^{T}CN_{\tau }+N_{\tau }^{T}CN_{\mu })]\overline{\Sigma }^{\prime }
\end{array}
\right.  \nonumber \\
&&  \label{ppp}
\end{eqnarray}
Writing 
\begin{eqnarray*}
&<&\phi ^{1}>=v_{1,\text{ }}<\phi ^{2}>=v_{2}, \\
&<&\Sigma >=\Lambda ,\text{ }<\Sigma ^{\prime }>=\Lambda ^{\prime }\text{ }
\end{eqnarray*}
the Dirac and Majorana mass matrices are given as 
\begin{equation}
m_{D}=\left( 
\begin{array}{lll}
h_{11}v_{2} & 0 & 0 \\ 
0 & h_{22}v_{1} & h_{22}v_{1} \\ 
0 & h_{32}v_{1} & h_{32}v_{1}
\end{array}
\right)  \label{p5}
\end{equation}
\begin{equation}
M_{R}=\left( 
\begin{array}{lll}
f_{11}\Lambda ^{\prime } & f_{12}\Lambda & f_{12}\Lambda \\ 
f_{12}\Lambda & f_{22}\Lambda ^{\prime } & f_{23}\Lambda ^{\prime } \\ 
f_{12}\Lambda & f_{23}\Lambda ^{\prime } & f_{22}\Lambda ^{\prime }
\end{array}
\right)  \label{p6}
\end{equation}
As is well known such a matrix is diagonalized by a mixing matrix with $\sin
^{2}\theta _{23}^{\prime }=\frac{1}{2}$ with $\theta _{13}^{\prime }=0$ i.e
by 
\begin{equation}
V=\left( 
\begin{array}{lll}
c^{\prime } & s^{\prime } & 0 \\ 
-\frac{s^{\prime }}{\sqrt{2}} & \frac{c^{\prime }}{\sqrt{2}} & \frac{1}{%
\sqrt{2}} \\ 
-\frac{s^{\prime }}{\sqrt{2}} & \frac{c^{\prime }}{\sqrt{2}} & -\frac{1}{%
\sqrt{2}}
\end{array}
\right) P(\gamma )  \label{p10}
\end{equation}
where $c^{\prime }=\cos \theta _{12}^{\prime },$ $s^{\prime }=\sin \theta
_{12}^{\prime }$ and $P(\gamma )$ is a diagonal phase matrix (consisting of
three non-trivial Majorana phases $\gamma _{1}$, $\gamma _{2},$ $\gamma _{3}$%
). It is now assumed that lepton number $L\equiv L_{e}$+$L_{\mu }$+$L_{\tau
} $ is non zero constant for $M_{R}$. This implies that [calculating $L$ for
the first row and second or third row of matrix $M_{R}$ and equating them] 
\begin{eqnarray*}
&&2f_{11}\Lambda ^{\prime }+2f_{12}\Lambda +f_{12}\Lambda +f_{12}\Lambda \\
&=&f_{12}\Lambda +f_{12}\Lambda +2f_{22}\Lambda ^{\prime }+f_{23}\Lambda
^{\prime }+f_{23}\Lambda ^{\prime }
\end{eqnarray*}
that is 
\begin{equation}
f_{11}\Lambda ^{\prime }+f_{12}\Lambda =f_{22}\Lambda ^{\prime
}+f_{23}\Lambda ^{\prime }  \label{new1}
\end{equation}
It has been pointed out [2] that this contraint implies tribimaximal mixing
matrix [3] i.e. 
\begin{equation}
\tan ^{2}\theta _{12}^{\prime }=\frac{1}{2}  \label{new2}
\end{equation}
Then 
\begin{equation}
V^{T}M_{R}V=\widehat{M}_{R}=diag(\widehat{M}_{1},\widehat{M}_{2},\widehat{M}%
_{3})  \label{p11}
\end{equation}
where 
\begin{eqnarray*}
\widehat{M}_{1} &=&M_{1}e^{2i\gamma _{1}}=[f_{11}\Lambda ^{\prime
}+2f_{12}\Lambda ]e^{2i\gamma _{1}} \\
\widehat{M}_{2} &=&M_{2}e^{2i\gamma _{2}}=[f_{11}\Lambda ^{\prime
}-f_{12}\Lambda ]e^{2i\gamma _{2}} \\
\widehat{M}_{3} &=&M_{3}e^{2i\gamma _{3}}=[f_{22}-f_{23}]\Lambda ^{\prime
}e^{2i\gamma _{3}}
\end{eqnarray*}

Then the effective Majorana mass matrix for light neutrinos is 
\begin{equation}
M_{\nu }=\widehat{m}_{D}\widehat{M}_{R}^{-1}\widehat{m}_{D}^{T}=\widehat{A}
\label{p15}
\end{equation}
where $\widehat{A}$ is $3\times 3$ matrix with matrix elements 
\begin{eqnarray}
a_{11} &=&h_{11}^{2}v_{2}^{2}A  \nonumber \\
\sqrt{2}a_{12} &=&h_{11}(2h_{22})v_{1}v_{2}B  \nonumber \\
\sqrt{2}a_{13} &=&h_{11}(2h_{32})v_{1}v_{2}B  \nonumber \\
a_{22} &=&\frac{1}{2}(4h_{22}^{2}v_{1}^{2})C  \label{p16} \\
a_{23} &=&\frac{1}{2}(2h_{22})(2h_{32})v_{1}^{2}C  \nonumber \\
a_{33} &=&\frac{1}{2}(4h_{32}^{2})v_{1}^{2}C  \nonumber
\end{eqnarray}
Here 
\begin{eqnarray}
A &=&e^{-2i\gamma _{1}}[\frac{c^{\prime 2}}{M_{1}}+\frac{s^{\prime 2}}{M_{2}}%
e^{2i(\gamma _{1}-\gamma _{2})}]  \nonumber \\
B &=&-e^{-2i\gamma _{1}}c^{\prime }s^{\prime }[\frac{1}{M_{1}}-\frac{1}{M_{2}%
}e^{2i(\gamma _{1}-\gamma _{2})}]  \nonumber \\
C &=&e^{-2i\gamma _{1}}[\frac{s^{\prime 2}}{M_{1}}+\frac{c^{\prime 2}}{M_{2}}%
e^{2i(\gamma _{1}-\gamma _{2})}]  \label{p16A}
\end{eqnarray}
In order to diagonalize $M_{\nu }$ as given in Eq. (\ref{p15}) I now make
the assumption of maximal atmospheric mixing and zero $U_{e3}$ or
equivalently $\mu -\tau $ symmetry for $M_{\nu }$ i.e. 
\begin{eqnarray*}
a_{12} &=&a_{13} \\
h_{22} &=&h_{33}\text{ }
\end{eqnarray*}
which also implies 
\begin{equation}
a_{23}=a_{33}  \label{pnew1}
\end{equation}
This requires the $3\times 3$ unitary matrix for diagonalization to be 
\begin{equation}
U=\left( 
\begin{array}{lll}
c & s & 0 \\ 
-\frac{s}{\sqrt{2}} & \frac{c}{\sqrt{2}} & \frac{1}{\sqrt{2}} \\ 
-\frac{s}{\sqrt{2}} & \frac{c}{\sqrt{2}} & -\frac{1}{\sqrt{2}}
\end{array}
\right) diag(e^{i\beta _{1}},\text{ }e^{i\beta _{3}},\text{ }e^{i\beta _{3}})
\label{p17A}
\end{equation}
so that 
\begin{equation}
U^{T}M_{\nu }U=diag(m_{1}\text{ }m_{2}\text{ }m_{3})  \label{p19}
\end{equation}

Finally then we have 
\begin{equation}
M_{\nu }=\left( 
\begin{array}{lll}
a & b & b \\ 
b & d & d \\ 
b & d & d
\end{array}
\right)  \label{p17}
\end{equation}
where 
\begin{eqnarray}
a &\equiv &a_{11}=e^{-2i\beta _{1}}[c^{2}m_{1}+s^{2}m_{2}e^{i\Delta }] 
\nonumber \\
2d &=&e^{-2i\beta _{1}}[s^{2}m_{1}+c^{2}m_{2}e^{i\Delta }]  \nonumber \\
\sqrt{2}b &=&-cse^{-2i\beta _{1}}[m_{1}-m_{2}e^{i\Delta }]  \label{p20} \\
m_{3} &=&0  \nonumber \\
\Delta &=&2(\beta _{1}-\beta _{2})  \nonumber
\end{eqnarray}
We wish to emphasize that form (\ref{p17}) for $M_{\nu }$ is a consequence
of $\mu -\tau $ symmetry for right handed $SU_{L}(2)$-singlet neutrinos, and
maximal atmospheric mixing and vanishing of $U_{e3}$. For tribimaximal
mixing, it is required that $a+b=2d$ or $a_{11}+a_{12}=a_{22}+a_{23}$ Using
Eqs. (\ref{p16}), (\ref{p16A}) and (\ref{pnew1}) it is easy to see that this
constraint requires that 
\begin{equation}
(h_{11}v_{2})^{2}=(2h_{22}v_{1})^{2}  \label{pnew2}
\end{equation}

On the other hand it has been shown in [1] that 
\begin{eqnarray}
\left| h_{11}v_{2}\right| ^{2}
&=&[(c^{2}m_{1}+s^{2}m_{2})^{2}-4c^{2}s^{2}m_{1}m_{2}\sin ^{2}\frac{\Delta }{%
2}]^{\frac{1}{2}}\times  \label{p27} \\
&&\{\frac{c^{\prime 2}}{M_{1}^{2}}+\frac{s^{\prime 2}}{M_{2}^{2}}-\frac{%
c^{2}s^{2}}{M_{1}M_{2}m_{1}m_{2}}[(m_{2}-m_{1})^{2}+4m_{1}m_{2}\sin ^{2}%
\frac{\Delta }{2}]\}^{-\frac{1}{2}}  \nonumber
\end{eqnarray}
\begin{eqnarray}
\left| 2h_{22}v_{1}\right| ^{2}
&=&m_{1}m_{2}[(c^{2}m_{1}+s^{2}m_{2})^{2}-4c^{2}s^{2}m_{1}m_{2}\sin ^{2}%
\frac{\Delta }{2}]^{-\frac{1}{2}}\times  \label{p28} \\
&&\{c^{\prime 2}M_{2}^{2}+s^{\prime 2}M_{1}^{2}-c^{2}s^{2}\frac{M_{1}M_{2}}{%
m_{1}m_{2}}[(m_{2}-m_{1})^{2}+4m_{1}m_{2}\sin ^{2}\frac{\Delta }{2}]\}^{%
\frac{1}{2}}  \nonumber
\end{eqnarray}

We now discuss leptogenesis in our scienario. As is well known [4,5,6,7] the
leptogenesis asymmetry is given by the matrix [12] 
\begin{equation}
\epsilon _{i}=\frac{1}{8\pi }\sum_{k\neq i}\frac{1}{v_{i}^{2}R_{ii}}\func{Im%
}[(R_{ik})^{2}f(\frac{M_{k}^{2}}{M_{i}^{2}})]  \label{p29}
\end{equation}
where $M_{i}$ denotes the heavy Majorana neutrino masses, $R_{ij}$ are
defined by the matrix [5] 
\begin{eqnarray}
R &=&\widehat{m}_{D}^{\dagger }\widehat{m}_{D}  \nonumber \\
&=&V^{T}m_{D}^{\dagger }m_{D}V^{*}  \label{pnew3}
\end{eqnarray}
The loop function f(x) containing vertex and self-energy corrections is 
\begin{equation}
f(x)=\sqrt{x}(\frac{2-x}{1-x}-(1+x)\ln \frac{1+x}{x})  \nonumber
\end{equation}

We have $\left| v_{1}\right| ^{2}+\left| v_{2}\right|
^{2}=(174GeV)^{2}=\left| v\right| ^{2}.$ We take $\left| v_{1}\right|
^{2}=\left| v_{2}\right| ^{2}=\frac{1}{2}v^{2},$ so that 
\begin{equation}
\epsilon _{1}=-\frac{1}{8\pi }f\left( \frac{M_{2}^{2}}{M_{1}^{2}}\right) 
\frac{1}{v_{1}^{2}R_{11}}\func{Im}[(R_{12})^{2}]  \label{p31}
\end{equation}
Using the constraint [4,5] 
\begin{equation}
R_{11}<4.3\times 10^{-7}v_{1}^{2},  \label{p36}
\end{equation}
obtained from out of equilibrium decay of $M_{1}\simeq 10^{10}GeV,$one
finally obtains from Eqs (\ref{pnew1}), (\ref{p31})and (\ref{p36}), the
lower limit on $\epsilon _{1}$ [1]: 
\begin{eqnarray}
\epsilon _{1} &=&-\frac{1}{8\pi }f\left( \frac{M_{2}^{2}}{M_{1}^{2}}\right) 
\frac{2.3\times 10^{6}}{v_{1}^{4}}c^{2}s^{2}\times  \nonumber \\
&&\times [\left| h_{11}v_{1}\right| ^{2}-\left| 2h_{22}v_{1}\right| ^{2}]^{2}%
\frac{(m_{2}^{2}-m_{1}^{2})M_{2}M_{1}}{(M_{2}^{2}-M_{1}^{2})m_{1}m_{2}}\sin
\Delta  \label{p37}
\end{eqnarray}
From it is clear that with the condition (\ref{pnew2}) , $\epsilon _{1}($ $%
\epsilon _{2})$ vanishes unless $M_{1}=M_{2}.$ For $M_{1}\simeq M_{2}$
writing 
\begin{eqnarray*}
M &=&\frac{M_{1}+M_{2}}{2},\text{ }\Delta M=\frac{M_{2}-M_{1}}{2}, \\
m &=&\frac{m_{1}+m_{2}}{2},\text{ }\Delta m=\frac{m_{2}-m_{1}}{2}, \\
(m_{2}^{2}-m_{1}^{2}) &=&4m\Delta m=\Delta _{solar}^{2}m
\end{eqnarray*}
we have 
\begin{equation}
f(x)=-\frac{M}{4\Delta M}  \label{pnew4}
\end{equation}
and from Eqs(\ref{p27}) and (\ref{p28}), in the leading order 
\begin{equation}
(\left| h_{11}v_{1}\right| ^{2}-\left| 2h_{22}v_{1}\right| ^{2})^{2}=\frac{4%
}{9}m^{2}M^{2}[(\frac{\Delta M}{M})+(\frac{\Delta m}{m})]^{2}  \label{pnew5}
\end{equation}
where we have used tribimaximal values $c^{2}-s^{2}=c^{\prime 2}-s^{\prime
2}=\frac{4}{9}.$ Then from Eq. (\ref{p37}) 
\[
\epsilon _{1}\simeq 6\times 10^{2}\frac{M^{2}}{v_{1}^{4}}\Delta
m_{solar}^{2}\sin \Delta \{1+\frac{\frac{\Delta m}{m}}{\frac{\Delta M}{M}}%
\}^{2} 
\]
The Majorana phase $\Delta $ is unknown, but is the same as would appear in
double beta decay [c.f. first of Eq. (\ref{p20})]. With $\sin \Delta \simeq
0.14$\cite{4}, and $\Delta m_{solar}^{2}=8\times 10^{-5}eV^{2},$ we can
write 
\begin{equation}
\epsilon _{1}\simeq 2\times 10^{-10}(\frac{M_{1}}{10^{10}GeV})^{2}\frac{%
\Delta m_{solar}^{2}}{8\times 10^{-5}eV^{2}}(\frac{174GeV}{v_{1}})^{4}\frac{%
\sin \Delta }{0.14}\{1+\frac{\frac{\Delta m}{m}}{\frac{\Delta M}{M}}\}^{2}
\label{p41}
\end{equation}

With $v_{1}^{2}=\frac{1}{2}v^{2}=\frac{1}{2}(174GeV)^{2}$, we get 
\begin{equation}
\epsilon _{1}\simeq 10^{-9}\{1+\frac{\frac{\Delta m}{m}}{\frac{\Delta M}{M}}%
\}^{2}  \label{p42}
\end{equation}

Finally using the neutrino oscillation data , we have 
\begin{equation}
m\simeq (\Delta m_{atm}^{2})^{\frac{1}{2}}=4.7\times 10^{-2}eV  \label{p44}
\end{equation}
\begin{equation}
\Delta m=\frac{\Delta m_{solar}^{2}}{4(\Delta m_{atm}^{2})^{\frac{1}{2}}}%
=4.3\times 10^{-4}eV  \label{p45}
\end{equation}
so that $\frac{\Delta m}{m}\simeq 10^{-2}$. Hence leptogenesis asymmtry $%
\epsilon _{1}$ as given in Eq. (\ref{p42}) is of the right order of
magnitude ($10^{-7}-10^{-6}$) providedd that $\frac{\Delta M}{M}\simeq
(1-3)\times 10^{-3}\left( \text{or}\frac{\Delta m}{m}/\text{ }\frac{\Delta M%
}{M}\simeq 10-30\right) $giving the level of degeneracy required for heavy
right handed neutrinos . It is important to note that CP violation
resposible for the generation of baryogenesis parameter through leptogenesis
comes entirely from Majorana phase $\Delta .$

One final comment is that although we have used the model $SU_{L}(2)\times
U_{e}(1)\times U_{\mu -\tau }(1)$, as a guide but the results are
independent of the details of this model.

The author would like to thank Prof. K. Sreenivasan for hospitality at Abdus
Salam International Centre for Theoretical Physics, Trieste where a part of
this work was done.

\end{document}